\documentclass[doublecol]{epl2} 

\usepackage{amssymb,amsfonts,bm}
\usepackage{graphics,graphicx}

\usepackage{fontenc}
\usepackage{t1enc}
\usepackage{amsmath}
\usepackage{float}
\usepackage{setspace}
\usepackage{appendix}
\usepackage{multirow}
\usepackage{color}
\usepackage{graphics,psfrag}
\usepackage{graphicx,psfrag}
\usepackage{enumerate}
\usepackage[none]{hyphenat}

\DeclareMathOperator{\Tr}{Tr}

\title{Critical properties of the eight-vertex model in a field}
\author{Roman Kr\v{c}m\'ar and Ladislav \v{S}amaj}
\shortauthor{R. Kr\v{c}m\'ar and L. \v{S}amaj}

\institute{Institute of Physics, Slovak Academy of Sciences, 
D\'ubravsk\'a cesta 9, SK-845 11, Bratislava, Slovakia, EU}

\pacs{64.60.F-}{Equilibrium properties near critical points, critical exponents}
\pacs{05.50.+q}{Lattice theory and statistics}
\pacs{05.70.Jk}{Critical point phenomena}

\abstract{The general eight-vertex model on a square lattice is studied 
numerically by using the Corner Transfer Matrix Renormalization Group method.
The method is tested on the symmetric (zero-field) version of the model, 
the obtained dependence of critical exponents on model's parameters is 
in agreement with Baxter's exact solution and weak universality is verified
with a high accuracy. 
It was suggested longtime ago that the symmetric eight-vertex model is 
a special exceptional case and in the presence of external fields 
the eight-vertex model falls into the Ising universality class. 
We confirm numerically this conjecture in a subspace of vertex weights, 
except for two specific combinations of vertical and horizontal fields 
for which the system still exhibits weak universality.}

\begin{document}

\maketitle

\section{Introduction}
The universality hypothesis states that for a statistical system with 
a given symmetry of microscopic state variables, critical exponents do
not depend on model's Hamiltonian parameters \cite{Griffiths70}.
Historically, the first violation of the universality was observed in
the symmetric (zero-field) eight-vertex model on the square lattice, 
whose critical exponents depend continuously on model's parameters.
Baxter solved the symmetric eight-vertex model by using the concept of 
commuting transfer matrices and the Yang-Baxter equation for the scattering 
matrix as the consistency condition 
\cite{Baxter71,Baxter72a,Baxter72b,Baxterbook}. 
This became a basis for generating and solving systematically integrable 
models within the ``Quantum Inverse-Scattering method''
\cite{Sklyanin78a,Sklyanin78b}, see e.g. monographs 
\cite{Korepinbook,Samajbook}. 
The next nonuniversal model, the Ashkin-Teller model 
\cite{Ashkin43,Fan72,Kadanoff77,Zisook80}, is in fact related to 
the eight-vertex model \cite{Kadanoff79}.
All these systems exhibit a ``weak universality'' as was proposed by 
Suzuki \cite{Suzuki74}: defining the singularities of statistical 
quantities near the critical point in terms of the inverse correlation
length, rather than the temperature difference, the rescaled 
critical exponents are universal.
The phenomenon of weak universality appears in many other physical systems, 
like interacting dimers \cite{Alet05}, frustrated spins \cite{Jin12}, 
quantum phase transitions \cite{Suzuki15} and so on.
There are indications that both universality and weak universality
are violated in the symmetric 16-vertex model on the 2D square 
and 3D diamond lattices \cite{Kolesik93a,Kolesik93b}, Ising spin glasses 
\cite{Bernardi95}, frustrated spin models \cite{Bekhechi03}, experimental
measurements on composite materials \cite{Omerzu01,Kagawa05}, etc.  

The general eight-vertex model on a square lattice can be formulated
as an Ising model on the dual square lattice with (nearest-neighbour
and diagonal) two-spin and (plaquette) four-spin interactions 
\cite{Wu71,Kadanoff71}. 
The symmetric version of the eight-vertex model corresponds to two
Ising models on two alternating sublattices, coupled with one another
via plaquette couplings. 
Kadanoff and Wegner \cite{Kadanoff71} suggested that the variation of
critical indices is due to the special hidden symmetries of the zero-field 
eight-vertex model. 
If an external field is applied, they argued that the magnetic exponents 
should be constant and equivalent to those of the standard Ising model,
see also monograph \cite{Baxterbook}.
This conjecture was supported by renormalization group calculations
\cite{Leeuwen75,Kadanoff79,Knops80}.

Since the eight-vertex model in a field is non-integrable, the above 
conjecture about the Ising-type universality must be checked numerically.
To our knowledge, no numerical test was done in the past, probably because 
of high demands on numerical precision. 
In this letter, in order to achieve a very high accuracy, we apply
the Corner Transfer Matrix Renormalization Group (CTMRG) method,
having its origin in the renormalization of the density matrix
\cite{White92,White93,Schollwock05,Krcmar15}.
A subspace of vertex weights is chosen to ensure the symmetricity
of the corner transfer matrix \cite{Baxterbook}. 
The CTMRG method is first tested on the zero-field version of 
the eight-vertex model, the obtained dependence of critical exponents on 
model's parameters is in good agreement with Baxter's exact solution and 
weak universality is verified. 
In the presence of external fields, the critical indices of the
eight-vertex model turn out to be constant, equivalent to the Ising ones,
except for two specific combinations of vertical and horizontal fields 
for which the system still exhibits weak universality with critical indices
dependent on model's parameters.

\section{Model} 
\begin{figure}
  \centering
  \includegraphics[width=0.4\textwidth,clip]{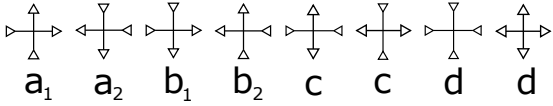}
  \caption{Admissible configurations of the eight-vertex model.}
  \label{obr:vertexy1}
\end{figure}

In vertex models, local state variables are localized on the edges of 
a lattice.
For each configuration of edge variables incident to a vertex, we associate
a Boltzmann weight.
For a given configuration of all edge states on the lattice, the contribution
to the partition function is the product of all vertex Boltzmann weights.
In the eight-vertex model on the square lattice, we have two-state 
arrow (dipole) edge variables. 
Each vertex satisfies the rule that only even number (i.e. 0, 2 or 4)
of arrows point toward it.
From among $2^4=16$ possible configurations 8 ones fulfill this rule,
see Fig. \ref{obr:vertexy1}.
Denoting by $E$ and $E'$ vertical and horizontal electric fields, respectively,
and by $T$ the temperature (in units of $k_{\rm B}=1$), the corresponding 
Boltzmann weights can be expressed as
\begin{equation} \label{fieldrepr}
\begin{split}
a_1 = & \, \, C \exp\left[ -\left( \epsilon_a-E-E' \right)/T \right] , \\
a_2 = & \, \, C \exp\left[ -\left( \epsilon_a+E+E' \right)/T \right] , \\
b_1 = & \, \, C \exp\left[ -\left( \epsilon_b+E-E' \right)/T \right] , \\
b_2 = & \, \, C \exp\left[ -\left( \epsilon_b-E+E' \right)/T \right] , \\
c = & \, \, C \exp\left( -\epsilon_c/T \right) , \\
d = & \, \, C \exp\left( -\epsilon_d/T \right) .
\end{split}
\end{equation}
Here, $\epsilon_a, \epsilon_b, \epsilon_c, \epsilon_d$ are energies invariant 
with respect to the reversal of all arrows incident to a vertex and the value
of the constant $C$ is irrelevant.

\begin{figure}
\centering
\includegraphics[width=0.1\textwidth,clip]{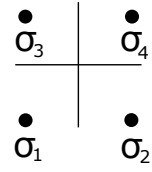}
\caption{Transformation from electric to magnetic Ising formulation.}
\label{obr:prepis}
\end{figure}

The eight-vertex model can be mapped into its Ising counterpart
defined on the dual (also square) lattice \cite{Wu71,Kadanoff71},
when assigning $+1$ to the arrows $\uparrow,\rightarrow$ and
$-1$ to the opposite arrows $\downarrow,\leftarrow$.
The Ising Hamiltonian can be written as $H = \sum_{\rm plaq} H_{\rm plaq}$,
where each square plaquette Hamiltonian $H_{\rm plaq}$ involves interactions
of four spins $\sigma_1,\sigma_2,\sigma_3,\sigma_4=\pm 1$ as depicted in 
Fig. \ref{obr:prepis}. 
Namely, we have horizontal nearest-neighbour interaction $J_h$ between 
$\sigma_1,\sigma_2$ and $\sigma_3,\sigma_4$, vertical nearest-neighbour
interaction $J_v$ between $\sigma_1,\sigma_3$, and $\sigma_2,\sigma_4$, 
diagonal interactions $J$ between $\sigma_1,\sigma_4$ and $J'$ between 
$\sigma_2,\sigma_3$ and finally four-spin interaction $J''$ between all spins
$\sigma_1,\sigma_2,\sigma_3,\sigma_4$, i.e.
\begin{eqnarray}
- H_{\rm plaq} & = & \frac{J_h}{2} (\sigma_1\sigma_2 + \sigma_3\sigma_4) +
\frac{J_v}{2} (\sigma_1\sigma_3 + \sigma_2\sigma_4) \nonumber \\
& & + J\sigma_1\sigma_4 + J'\sigma_2\sigma_3 
+J''\sigma_1\sigma_2\sigma_3\sigma_4 . \label{eq:ising_ham}
\end{eqnarray}
Note that nearest-neighbour couplings $J_h$ and $J_v$ are shared by
two plaquettes.
In terms of the Ising couplings, the original Boltzmann weights are written as
\begin{equation} \label{eq:magneticrepre1}
\begin{split}
a_1 = & \, \, C \exp\left[ \left( J_h+J_v+J+J'+J''\right)/T \right] , \\
a_2 = & \, \, C \exp\left[ \left( -J_h-J_v+J+J'+J''\right)/T \right] , \\
b_1 = & \, \, C \exp\left[ \left( J_h-J_v-J-J'+J''\right)/T \right] , \\
b_2 = & \, \, C \exp\left[ \left( -J_h+J_v-J-J'+J''\right)/T \right] , \\
c = & \, \, C \exp\left[ \left( -J+J'-J'' \right)/T \right] , \\
d = & \, \, C \exp\left[ \left( J-J'-J'' \right)/T \right] .
\end{split}
\end{equation}

The symmetric eight-vertex model corresponds to the case with no electric 
fields, $E=E'=0$.
Comparing (\ref{fieldrepr}) with (\ref{eq:magneticrepre1}) we see that
the horizontal and vertical nearest-neighbour Ising couplings vanish,
$J_h = J_v = 0$.
The system is thus composed of two alternating Ising sublattices, one with 
the two-spin coupling $J$ and the other with $J'$, the interaction between 
the sublattices being provided exclusively by the plaquette four-spin 
interactions $J''$. 
If $J''=0$, the system splits into two separated Ising lattices. 
The vertex weights (\ref{eq:magneticrepre1}) reduce themselves to
\begin{equation} \label{eq:magneticrepre2}
\begin{split}
a_1 = a_2 \equiv a , & \quad 
a = C \exp\left[ \left( J+J'+J''\right)/T \right] , \\
b_1 = b_2 \equiv b , & \quad 
b = C \exp\left[ \left( -J-J'+J''\right)/T \right] , \\
c = & \, \, C \exp\left[ \left( -J+J'-J'' \right)/T \right] , \\
d = & \, \, C \exp\left[ \left( J-J'-J'' \right)/T \right] .
\end{split}
\end{equation}

The symmetric eight-vertex model has five phases \cite{Baxterbook}.
We shall concentrate on the ferroelectric-A phase, defined by the inequality
$a > b + c + d$, and the disordered phase, defined by 
$a,b,c,d < (a + b + c + d)/2$.
The second-order transition between these phases takes place at 
the hypersurface 
\begin{equation} \label{crit}
a_c=b_c+c_c+d_c ,
\end{equation}
where $c$-subscript means evaluated at the critical temperature $T_c$.
In the special case $J''=0$ and $J'=J$, the relation (\ref{crit}) implies 
the well known critical condition for the Ising model
$2J/T_c = \ln(1+\sqrt{2})$.
Within the framework of the Ising representation, the magnetic critical 
exponents $\alpha$, $\beta$, $\gamma$ and $\nu$, which describe the singular 
dependence of statistical quantities on the small temperature difference 
$\Delta T = T_c-T$, are expressible in terms of the auxiliary parameter
\begin{equation} \label{eq:mu}
\mu = 2 \arctan \left( \sqrt{\frac{a_cb_c}{c_cd_c}} \right)
= 2 \arctan \left( e^{2J''/T_c} \right)
\end{equation}
as follows \cite{Baxterbook}
\begin{equation} \label{eq:crit_ex}
\alpha = 2 - \frac{\pi}{\mu} , \quad
\beta = \frac{\pi}{16\mu} , \quad
\gamma = \frac{7\pi}{8\mu} , \quad 
\nu = \frac{\pi}{2\mu} . 
\end{equation}
If $J''=0$, we have $\mu=\pi/2$ and Eq. (\ref{eq:crit_ex}) gives 
the standard 2D Ising indices 
\begin{equation} \label{eq:crit_ex1}
\alpha_{\rm I} = 0 , \quad \beta_{\rm I} = \frac{1}{8} , \quad
\gamma_{\rm I} = \frac{7}{4} , \quad \nu_{\rm I} = 1 . 
\end{equation}

Suzuki \cite{Suzuki74} proposed to express the singular behaviour of
statistical quantities close to the critical point in terms of 
the inverse correlation length $\xi^{-1}\propto (T_c-T)^{\nu}$
$(T\to T_c^-)$, instead of the temperature difference $T_c - T$.
The new (rescaled) critical exponents 
\begin{equation} \label{eq:crit_del}
\hat{\phi} \equiv \frac{2-\alpha}{\nu} = 2 , \quad 
\hat{\beta} \equiv \frac{\beta}{\nu} = \frac{1}{8} , \quad 
\hat{\gamma} \equiv \frac{\gamma}{\nu} = \frac{7}{4} 
\end{equation}
become universal and belong to the Ising universality class.
The remaining two exponents $\delta$ and $\eta$, defined just at 
the critical point, are constant and have their 2D Ising values
\begin{equation} \label{deltaeta}
\delta = 15, \qquad \eta = \frac{1}{4} .
\end{equation}
The phenomenon is known as ``weak universality''.

%In a $d$-dimensional space, scaling hypothesis implies that
%the relations \cite{Baxterbook}
%\begin{equation}\label{eq:weak_eta}
%\hat{\phi} = d, \quad \hat{\beta} = \frac{d-2+\eta}{2} , \quad 
%\hat{\gamma} = 2-\eta 
%\end{equation}
%and
%\begin{equation}
%\delta = \frac{d+2-\eta}{d-2+\eta} .
%\end{equation}
%Then the necessary condition for standard (say Ising) or weak universality 
%is that the critical exponent $\eta$ be constant.

\section{Method} \label{method}
The CTMRG method \cite{Nishino96a,Nishino97} is based on Baxter's corner 
transfer matrices \cite{Baxterbook}. 
Each quadrant of the square lattice with dimension $L\times L$ is represented 
by one corner matrix $C$ and the partition function $\mathcal{Z} = \Tr C^4$. 
The density matrix is defined by $\rho = C^4$, so that $\mathcal{Z}=\Tr\rho$. 
The number of degrees of freedom grows exponentially with $L$ and 
the density matrix is used in the process of their reduction. 
Namely, degrees of freedom are iteratively projected to the space generated 
by the eigenvectors of the density matrix with largest eigenvalues. 
Dimension of the truncated space is denoted by the $D$; the larger 
the value of $D$ taken, the better precision of the results is attained. 
The fixed boundary conditions are used, each spin at the boundary is
set to value $\sigma = -1$.
This choice ensures a quicker convergence of the method in the ordered phase.

From a technical point of view, it is important that the density matrix 
$\rho$ be symmetric.
It turns out that the symmetricity of $\rho$ is ensured by the condition
\begin{equation} \label{cd}
c = d ,
\end{equation}
which corresponds, in the Ising representation (\ref{eq:magneticrepre1}), 
to the constraint $J=J'$.
The subspace of vertex weights (\ref{cd}) involves both cases without 
and with external fields.
This is why the restriction (\ref{cd}), considered throughout the whole work,
does not prevent us from studying the effect of fields on critical properties
of the eight-vertex model.
We shall focus on the critical exponents $\nu$, $\eta$, $\beta$ and 
the central charge $c$.

The critical exponent $\nu$ can be obtained from the dependence of the 
internal energy $U$ on the linear size of the system $L$ at the critical
point \cite{Nishino96b},
\begin{equation} \label{UL}
U(L) - U(\infty) \sim L^{1/\nu - 2} , \qquad T = T_c .
\end{equation}
The effective (i.e. $L$-dependent) exponent $\nu_{\rm eff}$ is calculated 
as the logarithmic derivative of the internal energy as follows
\begin{equation}\label{eq:nu0f}
\nu_{\rm eff} = \left[ 3 + \frac{\partial}{\partial\ln L}
\ln\left(\frac{\partial U}{\partial L}\right) \right]^{-1} .
\end{equation}
If $T\ne T_c$, the plot $\nu_{\rm eff}(L)$ either goes quickly to 0 or diverges
as $L$ increases.
This means that we can determine the critical temperature $T_c$ from
the requirement
\begin{equation} \label{stab}
\lim_{L\to\infty} \nu_{\rm eff}(L) \to \nu ,
\end{equation}
where $0<\nu<\infty$ is the critical exponent we are looking for. 

The critical index $\eta$ can be deduced from the $L$-dependence of 
the magnetization $M = \langle \sigma \rangle$ at the critical point 
\cite{Nishino96b}, 
\begin{equation} \label{eq:fss}
M \sim  L^{-\eta/2} , \qquad T = T_c .
\end{equation}
The effective exponent $\eta_{\rm eff}$ is calculated as a logarithmic 
derivative of magnetization
\begin{equation} \label{eq:eta0f}
\eta_{\rm eff}=-2\frac{\partial \ln(M)}{\partial \ln(L)} .
\end{equation}
As before, $\eta = \lim_{L\to\infty} \eta_{\rm eff}(L)$.

To calculate the critical exponent $\beta$, we make use of the $T$-dependence 
of the spontaneous magnetization $M$ close to the critical temperature $T_c$:
\begin{equation} \label{eq:magnet}
M \propto (T_c-T)^{\beta} \qquad \mbox{as $T\to T_c^-$.}
\end{equation}
The critical exponent $\beta$ is extracted via the logarithmic derivative
\begin{equation} \label{eq:beta0f}
\beta_{\rm eff} = \frac{\partial\ln(M)}{\partial\ln(T_c-T)} .
\end{equation}
In general, $\beta_{\rm eff}$ as a function of $T$ has one extreme (maximum) 
at $T^*$, decays slowly for $T<T^*$ and drops abruptly for $T^*<T<T_c$, 
since the CTMRG method is inaccurate close to $T_c$. 
This is why we take as the critical index $\beta$ the maximal value of 
$\beta_{\rm eff}$, $\beta = \beta_{\rm eff}(T^*)$.

Another important quantity is the von Neumann entropy, defined by
\begin{equation}
S_{\rm N} = - \Tr \rho \ln \rho .
\end{equation}
Close to a critical point, it behaves as \cite{Calabrese04,Ercolessi10} 
\begin{equation}
S_{\rm N} \sim \frac{c}{6} \ln \xi ,
\end{equation}
where $c$ is the central charge.
Consequently, $S_{\rm N}$ has a logarithmic divergence at the critical point.
We ignore this alternative way of determining $T_c$ since the previous 
determination of $T_c$ via the stability condition (\ref{stab}) with a finite 
value of $\nu$ requires less computation and leads to more accurate results.
At the critical point, $S_{\rm N}$ grows with the system size $L$ as follows
\begin{equation}
S_{\rm N} \sim \frac{c}{6} \ln L , \qquad T=T_c . 
\end{equation}
The effective central charge is given by
\begin{equation}
c_{\rm eff} = 6 \frac{\partial S_{\rm N}}{\partial \ln L} 
\end{equation}
and the central charge $c = \lim_{L\to\infty} c_{\rm eff}(L)$.
We recall that $c=1/2$ for the universal Ising model and $c=1$ for 
the weakly universal symmetric eight-vertex model \cite{Baxterbook}.

\section{Test on the symmetric eight-vertex model}
We first test the CTMRG method on the exactly solved symmetric eight-vertex 
model with vertex weights (\ref{eq:magneticrepre2}), $c=d$. 
Baxter's critical exponents are given by Eqs. (\ref{eq:mu}) and 
(\ref{eq:crit_ex}).
We parametrize the vertex weights in such a way that on the critical 
hypersurface (\ref{crit}) one has
\begin{equation}
a_c = 1\quad (\epsilon_a=0), \qquad c_c = \frac{1-b_c}{2} .
\end{equation}
The value of the critical temperature is fixed to $T_c = 1$.

\begin{figure}
\centering
\includegraphics[width=0.46\textwidth,clip]{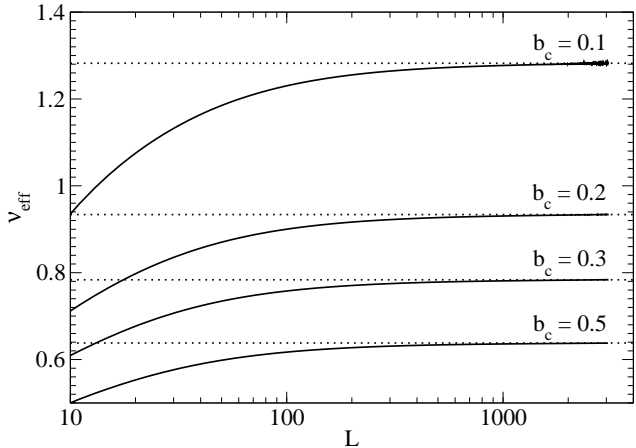}
\caption{The symmetric eight-vertex model: 
The dependence of the effective critical index $\nu_{\rm eff}$ on the system
size $L$, for four values of the critical vertex weight $b_c=0.1, 0.2, 0.3$ 
and $0.5$.
As $L$ increases, $\nu_{\rm eff}$ tends to the Baxter's exact value of $\nu$ 
represented by dotted line. $D=1000$.}
\label{obr:nun4}
\end{figure}

For four values of the critical vertex weight $b_c=0.1, 0.2, 0.3$ and $0.5$, 
the numerical results for the effective critical index $\nu_{\rm eff}$ as 
a function of the system size $L$ are pictured in Fig. \ref{obr:nun4}; 
hereinafter, the $L$-dependence of an effective critical index will be set 
in the logarithmic scale. 
It is seen that as $L$ increases $\nu_{\rm eff}$ tends to the Baxter's exact 
value of $\nu$ (horizontal dotted line). 

\begin{figure}
\centering
\includegraphics[width=0.48\textwidth,clip]{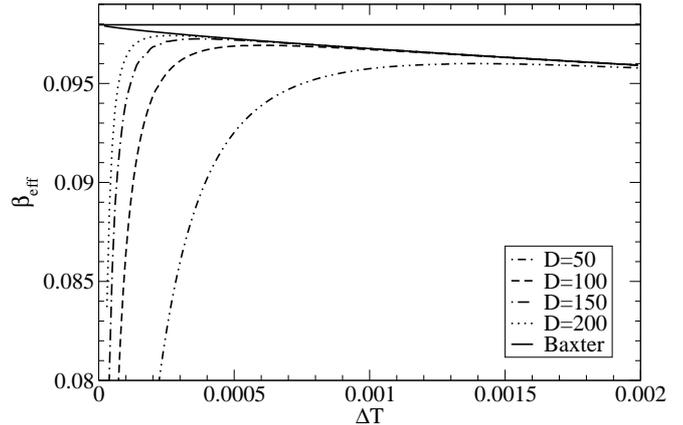}
\caption{The symmetric eight-vertex model with the critical vertex weight 
$b_c = 0.3$: 
The dependence of the effective critical index $\beta_{\rm eff}$ on the
distance from the critical temperature $\Delta T \equiv T_c-T$, for four
values of the truncation parameter $D=50, 100, 150$ and $200$. 
The exact Baxter result is represented by solid lines.}
\label{obr:beta03}
\end{figure}

\begin{figure}
\centering
\includegraphics[width=0.48\textwidth,clip]{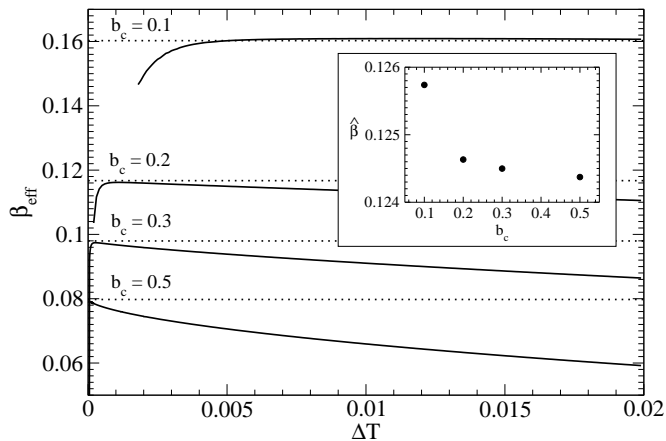}
\caption{The symmetric eight-vertex model:
The effective critical exponent $\beta_{\rm eff}$ as a function of 
$\Delta T$, for four values of the critical vertex weight 
$b_c=0.1, 0.2, 0.3, 0.5$ and the truncation order $D=200$.
The maxima of the plots are close to the Baxter exact results for $\beta$, 
represented by horizontal dotted lines. 
The inset shows an almost constant dependence of the rescaled critical index 
$\hat{\beta}\sim 1/8$ on $b_c$.}
\label{obr:betan4}
\end{figure}

The effective exponent $\beta_{\rm eff}$ is first plotted as a function of 
the distance from the critical temperature $\Delta T \equiv T_c-T$ for one 
fixed value of the critical vertex weight $b_c=0.3$ in Fig. \ref{obr:beta03}.
As the dimension of the truncated space of the density matrix $D$ increases 
from 50 up to 200, the maximum of the $\beta_{\rm eff}(\Delta T)$ plot 
approaches systematically to the Baxter exact result for $\beta$, 
represented by solid lines. 

For the fixed truncation order $D=200$ and four values of the critical vertex 
weight $b_c=0.1, 0.2, 0.3, 0.5$, the effective exponent $\beta_{\rm eff}$ 
as a function of $\Delta T$ is plotted in Fig. \ref{obr:betan4}.
The maxima of the $\beta_{\rm eff}(\Delta T)$ plots are close to 
the Baxter exact results for $\beta$, represented by dotted lines. 
The inset of Fig. \ref{obr:betan4} shows the dependence of the
rescaled critical index $\hat{\beta}\equiv \beta/\nu$ on $b_c$.
We see that $\hat{\beta}$ varies slightly between $0.124-0.126$,
i.e. the numerical results indicate with a high accuracy that $\hat{\beta}$ 
is a constant, close to the exact Baxter value $1/8$.
The major source of numerical errors in our calculations is 
the dimension of the truncated space $D$. 
We see that there is a dispersion of the results for different values of the 
parameter $b_c$, even though they are calculated with the same value of $D$. 
The dispersion originates from the fact, that each set of vertex parameters 
represents a different system with different rate of convergence. 
We can expect a comparable dispersion of values of the exponent $\hat{\beta}$ 
for the eight-vertex model with fields.

\begin{figure}
\centering
\includegraphics[width=0.48\textwidth,clip]{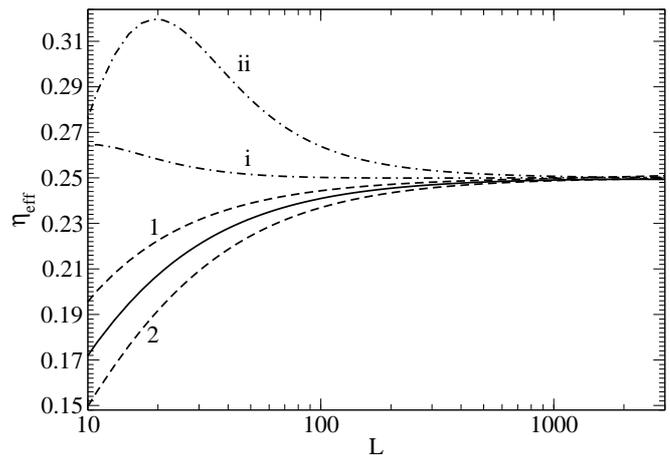}
\caption{The effective critical exponent $\eta_{\rm eff}$ as a function of 
the system size $L$.
The solid curve corresponds to the symmetric eight-vertex model with
$b_c=0.3$, the dashed lines 1 and 2 to the partially symmetric cases
(\ref{1}) and (\ref{2}), respectively, the dash-dotted lines i and ii to
the non-symmetric cases (\ref{i}) and (\ref{ii}), respectively.
In all cases, as $L$ increases $\eta_{\rm eff}$ goes asymptotically to 
$\eta = 1/4$.
$D=1000$.} 
\label{obr:eta}
\end{figure}

\begin{figure}
\centering
\includegraphics[width=0.48\textwidth,clip]{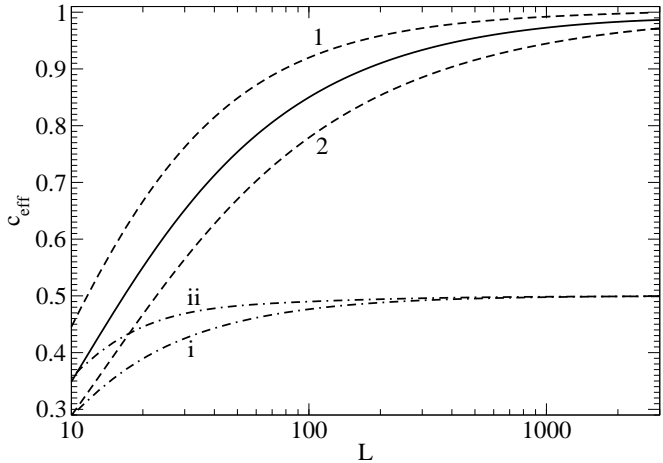}
\caption{The effective central charge $c_{\rm eff}$ as a function of 
the system size $L$.
Notation of curves as in Fig. \ref{obr:eta}.
As $L$ increases, the symmetric and partially symmetric eight-vertex models 
tend to $c=1$, the non-symmetric models to the Ising $c=1/2$.
$D=1000$.} 
\label{obr:central4}
\end{figure}

For the symmetric eight-vertex model with $b_c=0.3$, the solid curve in 
Fig. \ref{obr:eta} shows the size $L$-dependence of the effective critical 
exponent $\eta_{\rm eff}$. 
The curve converges to the Ising value $\eta=1/4$ as it should be. 
The effective central charge $c_{\rm eff}$ as a function of $L$ is pictured 
in Fig. \ref{obr:central4} by a solid line.
For large $L$, $c_{\rm eff}$ goes to $c=1$ which is the central charge
of the weakly universal symmetric eight-vertex model.

\section{The eight-vertex model in a field} \label{4spin}
For the eight-vertex model in a field, we distinguish between two cases.

In the partially symmetric case, we keep the symmetry of either $a$'s
or $b$'s vertex weights: 
\begin{equation} \label{a}
a_1 = a_2 = a , \qquad  b_1\ne b_2 ,
\end{equation}
or
\begin{equation} \label{b}
a_1 \ne a_2 , \qquad  b_1 = b_2 = b .
\end{equation}
As follows from the representation (\ref{fieldrepr}), the eight-vertex model
(\ref{a}) corresponds to nonzero external fields $E = -E'$ and 
the one (\ref{b}) to $E = E'$.
For simplicity, we shall concentrate on the version (\ref{a}) and consider
two specific choices of vertex weights, denoted as 1 and 2.
\begin{itemize}
\item
The choice 1 is characterized by $T_c = 0.512195$ and
\begin{equation} \label{1}
\begin{array}{ll}
a_c = 0.4828 , & \cr 
b_{1c} = 0.0546 , & b_{2c} = 0.1193 , \cr
c_c = 0.1974 . & 
\end{array}
\end{equation}
\item
The choice 2 is characterized by $T_c = 0.987774$ and
\begin{equation} \label{2}
\begin{array}{ll}
a_c = 1 , & \cr
b_{1c} = 0.3230 , & b_{2c} = 0.4843 , \cr  
c_c = 0.2956 . &
\end{array}
\end{equation}
\end{itemize}

In the non-symmetric case, both vertex weights $a$'s and $b$'s are unequal: 
\begin{equation} \label{ab}
a_1 \ne a_2 , \qquad  b_1\ne b_2 .
\end{equation}
The non-symmetric eight-vertex model corresponds to nonzero external fields
$E$ and $E'$, such that $E\ne \pm E'$.
We consider two choices of vertex weights, denoted as i and ii.
\begin{itemize}
\item
The choice i is characterized by $T_c = 0.740096$ and
\begin{equation} \label{i}
\begin{array}{ll}
a_{1c} = 0.6916 , & a_{2c} = 0.5278 , \cr
b_{1c} = 0.1530 , & b_{2c} = 0.2005 , \cr 
c_c = 0.3253 . &
\end{array}
\end{equation}
\item
The choice ii is characterized by $T_c = 1.172793$ and
\begin{equation} \label{ii}
\begin{array}{ll}
a_{1c} = 1.0890 , & a_{2c} = 0.9183 , \cr
b_{1c} = 0.4204 , & b_{2c} = 0.49856 , \cr 
c_c = 0.3582 . &
\end{array}
\end{equation}
\end{itemize}

\begin{figure}
\centering
\includegraphics[width=0.46\textwidth,clip]{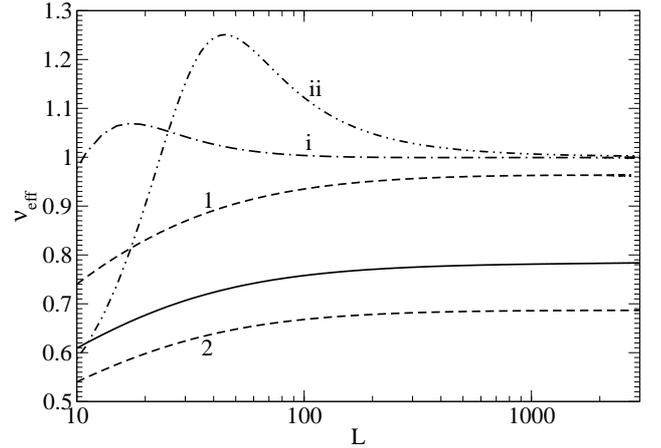}
\caption{The effective critical exponent $\nu_{\rm eff}$ as a function of
the system size $L$.
Notation of lines as in Fig. \ref{obr:eta}.
$D=1000$.} 
\label{obr:nu4}
\end{figure}

The numerical results for the effective critical index $\nu_{\rm eff}$ as 
a function of the system size $L$ are presented in Fig. \ref{obr:nu4}.
It is seen that for the symmetric eight-vertex model with $b_c=0.3$ 
(solid curve) as well as for the partially symmetric cases 
(\ref{1}) and (\ref{2}) (dashed lines 1 and 2, respectively),
as $L$ increases $\nu_{\rm eff}$ tend to parameter's dependent values
of $\nu$.
On the other hand, for both non-symmetric cases (\ref{i}) and (\ref{ii})  
represented by the dash-dotted lines i and ii, respectively, $\nu_{\rm eff}$
approaches to the Ising value of $\nu=1$.  

\begin{figure}
\centering
\includegraphics[width=0.48\textwidth,clip]{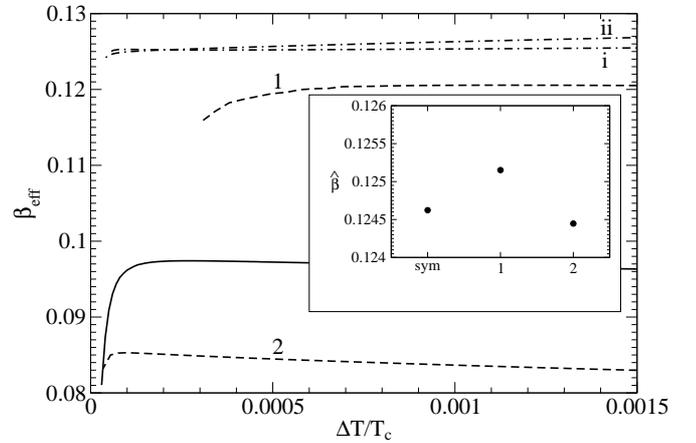}
\caption{The effective critical exponent $\beta_{\rm eff}$ as a function of
$\Delta T/T_c$.
Notation of curves as in Fig. \ref{obr:eta}.
The inset documents that the rescaled critical index 
$\hat{\beta}\sim 1/8$ for the symmetric and partially symmetric cases 1 and 2, 
confirming in this way their weak universality.
$D=200$ for the main figure and $D=300$ for the inset; the critical index
$\nu$ is calculated with $D=1500$.}
\label{obr:beta4}
\end{figure}

The dependence of the effective critical index $\beta_{\rm eff}$  
on the size $L$ is presented in Fig. \ref{obr:beta4}.
As before, for the partially symmetric cases 1 and 2, as $L$ increases 
the maxima of $\beta_{\rm eff}$ indicate parameter's dependent values of $\beta$.
We show in the inset that the rescaled critical index $\hat{\beta}\sim 1/8$ 
for these partially symmetric cases, confirming in this way their weak 
universality. 
We have chosen the range of $\hat{\beta}$-axis in between 
$0.124-0.126$, which is the anticipated dispersion of the weak-universality
results based on the numerical treatment of the exactly solvable symmetric case 
(see Fig. \ref{obr:betan4}).
For both non-symmetric cases i and ii, $\beta_{\rm eff}$ is consistent with 
the fixed Ising value of $\beta=1/8$.

\begin{figure}
\centering
\includegraphics[width=0.48\textwidth,clip]{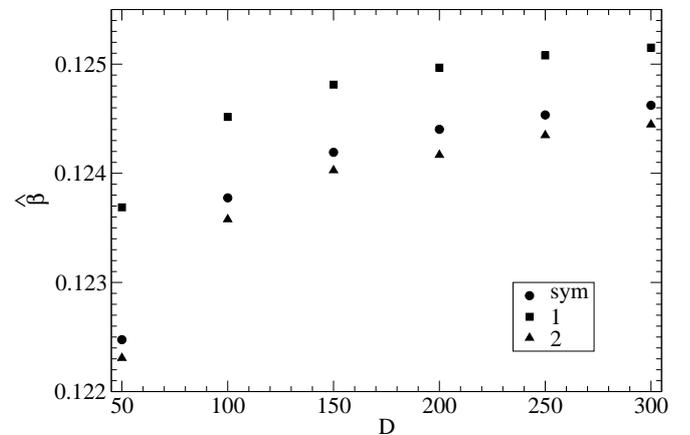}
\caption{
The rescaled critical index $\hat{\beta}=\beta/\nu$ as a function of the
dimension of the truncated space $D$, for the symmetric ($\bullet$)
and partially symmetric cases 1 ($\scriptstyle{\blacksquare}$) and 
2 ($\blacktriangle$).
The critical index $\nu$ is calculated with $D=1500$.}
\label{obr:error}
\end{figure}

In Fig. \ref{obr:error}, for the symmetric and partially symmetric 
1 and 2 cases, we present the convergence of the rescaled exponent 
$\hat{\beta}\equiv \beta/\nu$ as a function of the dimension of 
the density-matrix truncated space $D$ used for determining the exponent 
$\beta$; the dimension $D$ for the exponent $\nu$ is constant, $D=1500$.
As $D$ increases, the values of $\hat{\beta}$ approach the expected $1/8$.
Tiny deviations from $1/8$ are caused by the error in the determination
of the exponent $\nu$.

As concerns the effective critical exponent $\eta_{\rm eff}$, all curves
in Fig. \ref{obr:eta} converge as $L\to\infty$ to the same $\eta=1/4$. 

The effective central charge $c_{\rm eff}$ is presented as a function of 
size $L$ in Fig. \ref{obr:central4}.  
For the partially symmetric cases 1 and 2, as $L$ increases $c_{\rm eff}$ 
goes to $c=1$ which is the central charge of the weakly universal symmetric 
eight-vertex model.
For both non-symmetric cases i and ii, $c_{\rm eff}$ tends for large $L$ 
to $c=1/2$ which corresponds to the Ising universality class.

\section{Conclusion} \label{conclusion}
In this letter, we have studied the effect of external fields on critical
properties of the eight-vertex model on the square lattice. 
The model was studied numerically by using the CTMRG method which represents
a powerful mean to calculate accurately the critical temperature, critical 
exponents and the central charge $c$.
Within the magnetic representation of the eight-vertex model, we have
calculated the critical exponents $\nu$ and $\beta$, which are sufficient
to investigate the phenomenon of weak universality, and the exponent $\eta$, 
which is anticipated to be the same for all cases. 
The exactly solvable symmetric (zero-field) eight-vertex model exhibits
weak universality which was verified numerically with a high precision, 
see Figs. \ref{obr:nun4} and \ref{obr:betan4} with the inset.     
Kadanoff \cite{Kadanoff71} and Baxter \cite{Baxterbook} conjectured that 
the presence of nonzero external fields destroys weak universality and
the system belongs to the Ising universality class. 
We have checked numerically this conjecture in a subspace of vertex weights
(\ref{cd}) which ensures the symmetricity of the density matrix $\rho$. 
Our conclusion is that in the presence of fields one has to distinguish 
between the partially symmetric case, see Eqs. (\ref{a}) and (\ref{b}), and
the fully non-symmetric case (\ref{ab}).
The non-symmetric case, represented in Figs. \ref{obr:eta}-\ref{obr:beta4} 
by dash-dotted curves i and ii, evidently belongs to the Ising universality 
class with critical exponents independent of model's parameters and $c=1/2$, 
in agreement with the conjecture.
However, the partially symmetric case with nonzero fields $E$ and $E'$
such that $E=\pm E'$, represented in Figs. \ref{obr:eta}-\ref{obr:beta4} 
by dashed lines 1 and 2, has critical exponents $\nu$ and $\beta$ dependent
on model's parameters and exhibits weak universality (see the inset of
Fig. \ref{obr:beta4}) with $c=1$.
This contradicts Kadanoff's and Baxter's conjecture.

It would be interesting to extend the present treatment to the whole
space of vertex weights, without restriction (\ref{cd}).
This requires to diagonalize a non-symmetric density matrix which is
a nontrivial task.
The crucial question is whether the partially symmetric eight-vertex model
remains to be weakly universal when the $c=d$ symmetry is broken.
Another open question are the values of ``electric'' critical exponents
associated directly with the polarization and the arrow correlation 
function of the eight-vertex model. 

\begin{acknowledgments}
We are grateful to Andrej Gendiar for discussions about the CTMRG method.  
This work was supported by the project QETWORK APVV-14-0878 and VEGA Grants 
No. 2/0130/15 and No. 2/0015/15.
\end{acknowledgments}

\end{document}